\documentclass[12pt]{article}
\usepackage{epsf}
\usepackage{graphicx}
\usepackage{dcolumn}
\usepackage{bbm}
\usepackage{a4}
\usepackage{fancyhdr}
\usepackage{amsmath}
\usepackage{boxedminipage}
\usepackage{amssymb}
\usepackage{amsmath}
\usepackage{bbm}
\usepackage{graphics}
\usepackage{float}
\usepackage{psfrag}
\usepackage{epsfig}
\usepackage{graphicx}
\usepackage{tabularx}
\usepackage{calc}
\newcommand{\bi}{\bigskip}
\newcommand{\no}{\noindent}
\newcommand{\be}{\begin{eqnarray}}
\newcommand{\ee}{\end{eqnarray}}
\newcommand{\hk}{\hspace{0.1cm}}

\newcommand{\rk}{\right)}
\newcommand{\lk}{\left(}

\newcommand{\pli}{\prod\limits}
\newcommand{\il}{\int\limits}

\newcommand{\vk}{\vec{k}}
\newcommand{\vq}{\vec{q}}
\renewcommand{\vec}[1]{\mbox{\boldmath$#1$\unboldmath}}

\begin{document}

\title{The Yang-Mills Vacuum in Coulomb Gauge in $D=2+1$
Dimensions\thanks{Supported in part by DFG under Re 856/6-1 and Re 856/6-2 and
by the European Graduate School Basel-Graz-T\"ubingen}}
\date{\today}

\author{C. Feuchter and H. Reinhardt\\
Institut f\"ur Theoretische Physik\\
Auf der Morgenstelle 14\\
D-72076 T\"ubingen\\
Germany}
%



\maketitle

\begin{abstract}
The variational approach to the Hamilton formulation of Yang-Mills theory in
Coulomb gauge developed by the present authors previously is applied to
Yang-Mills theory in $2+1$ dimensions and is confronted with the existing
lattice data. 
We show that the resulting Dyson-Schwinger equations (DSE) yield consistent
solutions in $2 + 1$ dimensions 
only for infrared divergent ghost form factor and gluon energy. The
obtained numerical solutions of the DSE reproduce the
analytic infrared results and are in satisfactory agreement with the existing
lattice date in the whole 
momentum range.
\end{abstract}

\noindent
{\bf pacs:} {11.10.Ef, 12.38.Aw, 12.38.Cy, 12.38.Lg}

\no
\section{Introduction \label{sec1} }

Recently, there has been a renewed interest in the formulation of Yang-Mills
theory in Coulomb gauge, both in the continuum theory  \cite{R1,R2,R3,R4}
and on the lattice \cite{R5,R6,R6a,R6b,R8}. In the
continuum formulation the Hamilton approach turns out to be very appealing, in
particular, for the description of the confining properties of the theory
\cite{R3}. The reason is that in Coulomb gauge Gauss' law can be
explicitly resolved resulting in a static potential between color charges
\cite{R10}. Recently, several papers have been devoted to a variational solution
of the Yang-Mills Schr\"odinger equation in Coulomb gauge \cite{R2,R3,R4}. In
particular, in refs. \cite{R3,R4} the 
 present authors have developed a variational approach to
Yang-Mills theory in Coulomb gauge which 
properly includes the curvature of the space
of (transversal) gauge orbits. 
Using a Gaussian type of ansatz for the
Yang-Mills vacuum wave functional minimization of the energy density results in
a set of coupled Dyson-Schwinger equations which can be solved analytically in
the infrared \cite{R1,R7} and has been solved numerically in the full
momentum range \cite{R3,R17} in $D = 3 + 1$ dimensions. 
An infrared divergent gluon energy and a
linearly rising static quark potential was found \cite{R3,R17}, both
signaling confinement. Furthermore, within this approach the spatial 't Hooft
loop \cite{'t Hooft:1977hy,Reinhardt:2002mb} was calculated in the vacuum state and a perimeter law was
found \cite{Reinhardt:2007wh}, which is the behavior expected for this disorder parameter in
a confinement phase.  In the present paper we
apply this approach to $2+1$ dimensions and confront the results with the
existing lattice data \cite{R8}.

The $2+1$ dimensional Yang-Mills theory is interesting in several aspects: When
a Higgs field is included, it represents the high temperature limit of the $3+1$
dimensional Yang-Mills theory, thereby the temporal component of the gauge field
$A_0$ becomes the Higgs field in the dimensional reduced theory. In many
respects the $2+1$ dimensional theory is easier to treat than its $3+1$
dimensional counter part, in particular, much larger lattices can be afforded 
in
$2+1$ dimensions. This will be crucial for a comparison of the continuum results
with the lattice data, since the lattices affordable in $3+1$ dimensions are by
far too small to allow for a reliable extraction of the infrared properties of
the Greens functions \cite{Fischer:2002eq,Fischer:2007pf}. 
In addition, $2+1$ dimensional Yang-Mills theory
is super-renormalizable.

The balance of the paper is as follows: In section 2 we briefly summarize the
essential ingredients of the Hamilton approach to Yang-Mills theory in Coulomb
gauge \cite{R10} and of 
the variational solution of the corresponding Yang-Mills Schr\"odinger
equation \cite{R3}. 
In section 3 the Dyson-Schwinger equations resulting from minimizing
the energy density are studied in the ultraviolet and their renormalization is
carried out. Some exact statements on their solutions are given in
section 4, where we also solve these equations analytically in the infrared. 
In section 5 we present our numerical results and compare them with
the existing lattice data. A short summary and some concluding remarks are given
in section 6. 

\section{The Hamilton approach to Yang-Mills theory in Coulomb gauge \label{sec2}}

Canonical quantization of Yang-Mills theory is usually performed in Weyl gauge
$A_0 = 0$ to avoid the problem arising from the vanishing of the 
canonical momentum conjugate to $A_0$.\footnote{An alternative to the Weyl gauge is the light-cone
gauge $A_0 \pm A_1 = 0$.} The prize one pays by choosing Weyl gauge is that one
looses Gauss' law as equation of motion. To ensure gauge invariance one has to
impose Gauss' law as a constraint on the wave functional. Instead of using gauge
invariant wave functionals \cite{R11,R12,Rx}, 
it is simpler to explicitly resolve Gauss'
law by fixing the residual (time-independent) gauge invariance,
 and Coulomb gauge
$\partial_i A_i = 0$ is  a particularly convenient gauge for this 
purpose. After resolving Gauss' law in Coulomb
gauge the Yang-Mills Hamiltonian reads \cite{R10}
\begin{align}
H & =  \frac{1}{2} \int d^2 x \left[ {\cal{J}}^{- 1} [A^\perp] \Pi^{\perp a}_i ({\bf x}) 
{\cal{J}} [A^\perp] \Pi^{\perp a}_i
({\bf x})  +  B^a_i ({{\bf x}})^2 \right] \nonumber\\
&  + \frac{g^2}{2} \int d^2 x \int d^2 x' {\cal{J}}^{- 1} [A^\perp] \rho^a ({\bf x}) 
F^{a b} \lk {{\bf x}},
{{\bf x}'} \rk {\cal{J}} [A^\perp] \rho^b ({{\bf x}'}) \hk ,
\label{1}
\end{align}
where 
\begin{align}
\Pi^{\perp a}_i ({\bf x}) = t_{i k} ({\bf x}) \frac{\delta}{i \delta A^a_k ({\bf x})} \equiv
\frac{\delta}{i \delta A^{\perp a}_i ({\bf x})}
\label{2}
\end{align}
is the canonical momentum operator conjugate to the transverse gauge field
$A^\perp$. Furthermore 
\begin{align}
{\cal{J}} [A^\perp] = \det \lk - \partial_i \hat{D}_i (A^\perp) \rk \hk 
\label{3}
\end{align}
is the Faddeev-Popov determinant with $\hat{D} (A) 
= \partial + g \hat{A} \hk , \hk
\hat{A} = A^a \hat{T}_a \hk , \hk \lk \hat{T}_a \rk^{c b} = f^{c a b}$ being the
covariant derivative in the adjoint representation. Furthermore 
\begin{align}
B = \frac{i}{2 g} \epsilon^{i j} [D_i , D_j] \hk , \hk D_i = \partial_i + i g
A_i 
\label{4}
\end{align}
is the magnetic field, which is a scalar in $2+1$ dimensions
($\epsilon_{ij} = - \epsilon_{ji}, \epsilon_{12} = 1$). Finally
\begin{align}
\label{5a}
F^{a b} \lk {{\bf x}, {\bf x}'} \rk = \langle {{\bf x}} a | ( - \hat{D}_i 
\partial_i )^{- 1} ( - \partial^2) ( - \hat{D}_j  \partial_j)^{- 1} 
| {{\bf x}'} b \rangle \hk \hk 
\end{align}
is the non-Abelian Coulomb propagator which mediates a static interaction
between the color charge density of the gluons
\begin{align}
\rho^a ({{\bf x}}) = - \hat{A}^{\perp a b}_i ({{\bf x}}) \Pi^{\perp b}_i ({{\bf x}}) .
\label{7}
\end{align}
In the presence of external color charges, for example in the presence of
quarks, their charge density has to be added to the gluon charge density. 
The kinetic term in the Hamilton has the form of a (variational extension of
the) Laplace-Beltrami operator in curved space with the Faddeev-Popov
determinant (\ref{3}) 
corresponding to the determinant of the metric of the space of
transversal gauge orbits. The
Jacobian (\ref{3})
also enters the scalar product in the Hilbert space of the Yang-Mills
wave functionals in Coulomb gauge 
\begin{align}
\langle \Psi_1 | O | \Psi_2 \rangle = \int {\cal D} A^{\perp} {\cal{J}} [A^{\perp}] \Psi^*_1 
[A^{\perp}] O [A^{\perp},\Pi^{\perp }] \Psi_2 [A^{\perp}] \hk .
\label{8}
\end{align}

We will solve
 the Yang-Mills Schr\"odinger equation by the variational principle
\begin{align}
E = \langle \Psi | H | \Psi \rangle \to \mbox{min} 
\label{9}
\end{align}
using the following ansatz for the vacuum wave functional \cite{R3}, \cite{R4}
\begin{align}
\Psi [A^{\perp}] = \langle \Psi | A^{\perp} \rangle = 
{\cal N} {\cal J} [A^\perp]^{- \alpha} \exp \left[ - \frac{1}{2} \int d^2
x \int d^2 x' A^{\perp a}_i  ({\bf x}) \omega
 ({\bf x}, {\bf x}') A^{\perp a}_i  ({\bf x}')
\right] \hk ,
\label{11}
\end{align}
where $\omega (\vec{x}, \vec{x'})$ is the variational kernel, which by
translational and rotational invariance of the vacuum depends only on $| \vec{x}
- \vec{x'}|$, and, by the isotropy of color space, is independent of color. The
ansatz (\ref{11}) with $\alpha = \frac{1}{2}$ is motivated by the wave
functional of a particle moving in a s-state in a spherically symmetric
 potential. In
principle, $\alpha$ could be used as a variational parameter to minimize the
energy density. However, it turns out that up to two loops in the energy,
stationarity of the energy density 
with respect to $\omega$, i.e. $\delta E / \delta 
\omega = 0$, implies also stationarity with respect to $\alpha$, i.e. $d E / d
\alpha = 0$ \cite{R4}. 
Thus, we are free to choose $\alpha$ for convenience, and as in ref.
\cite{R3} we will choose  $\alpha = \frac{1}{2}$, which removes
the Faddeev-Popov determinant from the integration measure in eq. (\ref{8}).
Furthermore with the choice  $\alpha = \frac{1}{2}$ the gluon propagator is
given by
\begin{align}
\langle \Psi | A^{\perp}_i (x) A^{\perp}_j (y) | \Psi \rangle = \frac{1}{2} t_{i j} (x)
\omega^{- 1} (x, y) \hk ,
\label{12}
\end{align}
so that the Fourier transform $\omega (k)$ has the meaning of the gluon energy. 

The  calculation of the vacuum expectation value of the Coulomb Hamiltonian in
the state (\ref{11}) proceeds in the same way as in  3+1-dimensions and we just
quote the result. We find for the kinetic energy
\be
\label{1-49}
E_k & =  & \frac{N^2_C - 1}{4} \delta^{(2)} (0) \int d^2 k \frac{\left[\omega (\vk) -
\chi (\vk) \right]^2}{\omega (\vk)} ,
\ee
the potential energy
\be
E_p & = & \frac{N^2_C - 1}{4} \delta^{(2)} ({\bf 0}) \int d^2 k \frac{{\bf
k}^2}{\omega ({\bf k})} \nonumber\\
& & + \frac{N_C (N^2_C - 1)}{16} g^2 \delta^{(2)} ({\bf 0}) \int \frac{d^2 k
d^2 k'}{(2 \pi)^2 } \frac{1}{\omega ({\bf k}) \omega ({\bf k'})} 
\left(1 - \frac{({\bf k}{\bf k'})^2}{{\bf k}^2 {\bf k'}^2} \right) 
\ee
and for the Coulomb energy
\be
\label{1-50}
E_c & = & \frac{N_C (N^2_C - 1)}{16} \delta^{(2)} (0) \int 
\frac{d^2 k d^2 k'}{(2 \pi)^2} 
\frac{({\bf k}{\bf k'})^2}{{\bf k}^2 {\bf k'}^2} 
\frac{d (\vk - \vk')^2 f (\vk - \vk')}{(\vk - \vk')^2} \nonumber\\
& & \cdot \frac{\lk [ \omega (\vk) - \chi (\vk)] - [\omega (\vk') - \chi (\vk')]
\rk^2 }{\omega (\vk) \omega (\vk')} \hk ,
\ee
where $\delta^2 ({\bf 0}) = V / (2 \pi)^2$ with $V = \int d^2 x$ being the
2-dimensional spatial volume. 
Furthermore the quantity
\begin{align}
\chi ({\bf x},{\bf x}') = - \frac {1}{N_C^2 -1}
\left\langle \frac{\delta^2 \ln {\cal{J}}}{\delta A^{\perp a}_i ({{\bf x}}) \delta A^{\perp a}_i ({\bf x}')} 
\right\rangle_\Psi 
\label{20}
\end{align}
is referred to as ``curvature''
of the space of gauge orbits. Introducing  the ghost
propagator by
\begin{align}
G = \langle \Psi | \lk - \partial_i \hat{D}_i \rk^{- 1} | \Psi \rangle = 
(- \partial^2 )^{-1} \frac {d (- \partial^2)}{g} \hk ,
\label{14}
\end{align}
with $d (k)$ being the ghost form factor, the curvature (\ref{20}) 
can be expressed as
\begin{align}
{\chi}({\bf k}) = 
\frac{N_C}{2} \int \frac{d^2 q}{(2 \pi)^2}  
\lk 1 - \frac{({\bf k} {\bf q})^2}{{\bf k}^2 {\bf q}^2} \rk
\frac{{d} ({\bf k} - {\bf q}) {d} ({\bf q})}{({\bf k} - {\bf q})^2}  .
\label{22}
\end{align}
Finally, $f (k)$ denotes the Coulomb form factor, which is defined by 
\begin{align}
\langle \Psi | \lk - \partial_i \hat{D}_i \rk^{- 1} 
(- \partial^2 ) \lk - \partial_i \hat{D}_i \rk^{- 1}
| \Psi \rangle = 
G (- \partial^2 ) f G  \hk .
\label{17}
\end{align}
Minimization of the energy density  $(E_k + E_p + E_c) / V$ 
results in the so-called gap equation, which
can be cast into the form of a dispersion relation of a relativistic particle
(gluon) 
\begin{align}
\omega ({\bf k})^2 = {\bf k}^2 + \chi ({\bf k})^2   + I_\omega ({\bf k})
+ I^0_\omega \hk .
\label{27}
\end{align}
Here $I^0_\omega$ is an irrelevant constant which arises from the gluon 
tadpole. Furthermore, 
the quantity $I_\omega(k)$, which arises from the
expectation value of the Coulomb term (\ref{1-50}), 
can be expressed as
\begin{align}
I_\omega ({\bf k}) & = \frac{N_C}{2} \int \frac{d^2 q}{(2 \pi)^2} 
\frac{({\bf k}{\bf q})^2}{{\bf k}^2 {\bf q}^2}  
 \frac{d ({\bf k} -{\bf q})^2 f ({\bf k} - {\bf q})}{({\bf k} - {\bf q})^2} 
\nonumber\\
& \hspace{5cm} \cdot \frac{\left[ \omega ({\bf q}) - \chi ({\bf q}) + \chi
({\bf k}) \right]^2 - \omega ({\bf k})^2}{\omega ({\bf q})} \hk .
\label{28}
\end{align}

In principle $G$ or $d$ , $\chi$ and $f$ are defined uniquely once the trial
wave functional $\Psi (A)$ is fixed. However,
the exact evaluation of these expectation values (even with the above chosen
trial wave function) is not feasible and one has to resort to further
approximations. We shall adopt here the same approximation as in ref. \cite{R3}
 used in $3+1$ dimensions, approximating the full ghost-gluon
vertex by its bare one. This approximation has been justified in Landau gauge
\cite{R13} 
and, in fact, has received recently strong support by lattice calculations
\cite{Maas:2007uv}. In
this approximation the ghost and Coulomb form factors satisfy the following
Dyson-Schwinger equations
\begin{align}
\label{17a}
\frac{1}{{d} ({\bf k})}  
= 1 - {I}_d({\bf k}) \hk , 
\end{align}
\begin{align}
\label{18a}
{I}_d({\bf k}) = \frac{N_C}{2} \int \frac{d^2 {q}}{(2 \pi)^2}
\lk 1 - \frac{({\bf k} {\bf q})^2}{{\bf k}^2 {\bf q}^2} \rk
\frac{ {d} ({\bf k}  - {\bf q})}{({\bf k} - {\bf q} )^2 {\omega} ({\bf q})} ,
\end{align}

\begin{align}
{f} ({\bf k}) = 1 + {I}_f({\bf k}) ,
\label{18}
\end{align}

\begin{align}
{I}_f({\bf k}) = \frac{N_C}{2} \int \frac{d^2 q}{(2 \pi)^2} 
\lk 1 - \frac{({\bf k} {\bf q})^2}{{\bf k}^2 {\bf q}^2} \rk
\frac{{d} ({\bf k} - {\bf q})^2 {f} 
({\bf k} - {\bf q})}{({\bf k} -
{\bf q})^2 {\omega} ({\bf q})}  \hk .
\label{19}
\end{align}
As described in ref. \cite{R3} to the considered order (bare
ghost-gluon vertex and 2-loop order in energy) the Coulomb form factor can 
be put
$f (k) = 1$.

In $D = 2 + 1$ the coupling constant $g$ has the dimension of the square root
the momentum and it is
convenient to use the coupling constant to rescale all quantities by suitable
powers of $g$ to render them dimensionless. The coupling constant then
disappears from the Dyson-Schwinger equations.  Denoting the dimensionless
quantities by a bar we have
\be
\bar{k} & = & \frac{k}{g^2} \hk , \hk \bar{\omega} (\bar{k}) = 
\frac{\omega (g^2\bar{k})}{g^2} \hk , \hk 
\bar{\chi} (\bar{k}) = \frac{\chi  (g^2 \bar{k})}{g^2}
\nonumber\\
\bar{d} (\bar{k}) & = & \frac{d (g^2 \bar{k})}{g} \hk , \hk \bar{f}
(\bar{k})
= f (g^2 \bar{k}) \hk .
\ee
In the following we will skip the bar and, 
unless stated otherwise, all quantities will be understood as the dimensionless ones.

In section \ref{sec4.1} we prove that for any solution of the coupled Dyson-Schwinger
equations (\ref{27}, \ref{17a}) 
in $D = 2 + 1$ the ghost form factor and the gluon energy are
infrared divergent 
\be
\label{an4}
d^{- 1} (k = 0) & = & 0 \\
\label{an5}
\omega^{- 1} (k \to 0) & = & 0 \hk .
\ee
The first relation is the so-called horizon condition which had to be imposed ad
hoc in $D = 3 + 1$ but is a strict consequence of the Dyson-Schwinger equations
in $D = 2 + 1$. The second condition (\ref{an5}) signals gluon confinement.
In addition, we prove  in appendix A that when the curvature $\chi (k)$ is
omitted, the coupled Dyson-Schwinger equations in $D = 2 + 1$ do not allow for a
consistent solution. This is again different from $D = 3 + 1$ where the DSE
allow for solutions with infrared finite $\omega (k)$ when the curvature $\chi
(k)$ is ignored.

\section{Ultraviolet behavior and renormalization}

In the following we will investigate the ultraviolet behavior of
the solutions of the Dyson-Schwinger equations and perform their renormalization, 
thereby following the procedure
presented in refs. \cite{R3}, \cite{Reinhardt:2007wh} for the $D = 3 + 1$-dimensional case.
Since most of the considerations will parallel the $3 + 1$-dimensional case we will
be very brief.

For the uv-analysis it is sufficient to use the angular approximation
\be
\label{1-52}
h (|\vk - \vq|) = h (k) \Theta (k - q) + h (q) \Theta (q - k) \hk .
\ee
The angular integral in the ghost Dyson-Schwinger equation and also in the
curvature then becomes trivial 
$\il^{2 \pi}_0 d \varphi \sin^2 \varphi = \pi$ and 
we obtain for the corresponding momentum integrals 
\be
\label{1-54}
I_d (k) & = & \frac{N_C}{8 \pi} \left[ \frac{d (k)}{k^2} \il^k_0 d q
\frac{q}{\omega (q)} + \il^\infty_k d q \frac{d (q)}{q \omega (q)} \right] \\
\label{1-55}
\chi (k) & = & \frac{N_C}{8 \pi} \left[ \frac{d (k)}{k^2} \il^k_0 d q
q d (q) + \il^\Lambda_k d q \frac{d (q)^2}{q} \right] \hk .
\ee
The asymptotic analysis of the Dyson-Schwinger equations is simplified by taken
the derivative with respect to the external momentum
\be
\label{1-56}
I'_d (k) & = & \frac{N_C}{8 \pi} \frac{1}{k^2} \left[ d' (k) - 2 \frac{d (k)}{k}
\right] \il^k_0 d q \frac{q}{\omega (q)} \\
\label{1-57}
\chi' (k) & = & \frac{N_C}{8 \pi} \frac{1}{k^2} \left[ d' (k) - 2 \frac{d (k)}{k}
\right] \il^k_0 d q q d (q) \hk .
\ee
As we will see below, the remaining momentum integrals are ultraviolet
convergent\footnote{The ghost integral $I_d (k)$ is uv-finite as will be shown
later.}. 
The derivative of the ghost Dyson-Schwinger equation (\ref{17a}) yields
\be
\label{1-58}
d' (k) \left[ \frac{1}{d (k)^2} - \frac{N_C}{8 \pi} \frac{R (k)}{k^2} \right] =
- \frac{N_C}{4 \pi} \frac{d (k)}{k} \frac{R (k)}{k^2} \hk ,
\ee
where
\be
\label{1-59}
R (k) = \il^k_0 d q \frac{q}{\omega (q)} \hk 
\ee
and the derivative of the curvature (\ref{22}) yields
\be
\label{1-60}
\chi' (k) = \frac{N_C}{8 \pi} \frac{1}{k^2} \left[ d' (k) - 2 \frac{d (k)}{k}
\right] S (k) \hk ,
\ee
where
\be
\label{1-61}
S (k) = \il^k_0 d q q  d (q) \hk .
\ee
We now discuss the ultraviolet behavior of the relevant
quantities.

\subsection{Ultraviolet behavior}

For large momenta the $k^2$ term on the r.h.s. of the gap equation (\ref{27})
dominates \cite{R3} and  the gluon energy
$\omega (k)$ behaves like
\be
\label{1-62}
\omega (k) \to \sqrt{\vk^2} \hk , \hk k \to \infty \hk 
\ee
in accordance with asymptotic freedom. This behavior will in fact be confirmed by the numerical solutions presented in
section 6. Assuming (\ref{1-62}), we will investigate the uv-behavior of the
remaining quantities: the ghost form factor $d (k)$, the curvature $\chi (k)$
and the Coulomb form factor $f (k)$. 

For large $k \to \infty$ the $k$-dependence
of the integral (\ref{1-59}) is independent of the infrared behavior of $\omega
(k)$ and with eq. (\ref{1-62}) we find
\be
\label{1-63}
R (k) = k \hk , \hk k \to \infty \hk 
\ee
and the derivative of the ghost DSE (\ref{1-58}) reduces to
\be
\label{1-64}
\frac{d' (k)}{d (k)^2} \left[ 1 - \frac{N_C}{8 \pi} \frac{d (k)^2}{k} \right] =
\frac{N_C}{4 \pi} \frac{d (k)}{k^2} \hk .
\ee
To solve this equation, let us assume for the moment $\frac{d (k)^2}{k} \ll 1$
for $k \to \infty$. Then the differential equation (\ref{1-64}) reduces to
\be
\label{1-65}
\frac{d' (k)}{d (k)^2} = - \frac{N_C}{4 \pi} \frac{d (k)}{k^2} \hk ,
\ee
whose solution is given by
\be
\label{1-66}
d (k) = \frac{1}{\sqrt{\frac{1}{c^2} - \frac{N_C}{8 \pi} \frac{1}{k}}} \hk ,
\ee
where $c$ is an integration constant. Indeed, for large $k \to \infty$ this
solution satisfies $\frac{d (k)^2}{k} \ll 1$. To determine the integration
constant $c$, we consider the asymptotic behavior of the integral $I_d (k)$ 
(\ref{1-54}) for $k \to \infty$, where $\omega (k) \simeq k$ and $d
(k) \simeq c$. This yields
\be
\label{1-67}
I_d (k) \simeq \frac{N_C}{8 \pi} \left[ \frac{c}{k^2} \il^k_0 d q
\frac{q}{\omega (q)} + c \il^\infty_k d q \frac{1}{q \omega (q)} \right] \simeq
\frac{N_C}{4 \pi} \frac{c}{k} \hk .
\ee
Since $I_d (k \to \infty) \to 0$, we obtain from the DSE of the ghost form
factor $d (k \to \infty) \to 1$, which fixes the integration constant in eq.
(\ref{1-66}) to $c = 1$, so that the asymptotic form of the ghost form factor
becomes
\be
\label{1-68}
d (k) = \frac{1}{\sqrt{1 - \frac{N_C}{8 \pi} \frac{1}{k}}} \hk , \hk k  \to
\infty \hk .
\ee
Accordingly we find for the unscaled form factor\footnote{Below 
we denote the dimensionful (unscaled) quantities by a ''tilde''.}
$\tilde{d} (\tilde{k}) = g d
\lk \frac{\tilde{k}}{g^2} \rk $ the asymptotic form
\be
\label{1-69}
\tilde{d} (\tilde{k}) = \frac{g}{\sqrt{1 - \frac{N_C}{8 \pi}
\frac{g^2}{\tilde{k}}}} \hk , \hk k  \to
\infty \hk .
\ee
From the Swift relation \cite{R15}, \cite{R3} 
\be
\label{1-70}
\tilde{f} (\tilde{k}) = - \frac{1}{g^2} \frac{\partial}{\partial g}
\frac{1}{\tilde{d} (\tilde{k})}
\ee
we find for the Coulomb form factor
\be
\label{1-71}
f (k) = \frac{1}{\sqrt{1 - \frac{N_C}{8 \pi} \frac{1}{k}}} = d (k) \hk , \hk k
\to \infty \hk ,
\ee
which obviously has the same asymptotic form as the ghost form factor. 

With the asymptotic behavior of the ghost (\ref{1-68}), we find for the
derivative of the curvature (\ref{1-60})
\be
\label{1-73}
\chi' (k) = - \frac{N_C}{4 \pi} \frac{1}{k^3} S (k) \hk , \hk k \to \infty \hk 
\ee
and for the integral $S (k)$ (\ref{1-61})
\be
\label{1-75}
S (k \to \infty) \to \frac{k^2}{2} \hk ,
\ee
so that 
\be
\label{1-76}
\chi' (k) = \frac{N_C}{8 \pi} \frac{1}{k} \hk , \hk k \to \infty \hk ,
\ee
i.e.
\be
\label{1-77}
\chi (k) \sim \ln \lk \frac{k}{\mu} \rk \hk , \hk k \to \infty \hk .
\ee
Accordingly, the ratio
\be
\label{1-78}
\frac{\chi (k)}{\omega (k)} \sim \frac{1}{k} \ln \lk \frac{k}{\mu} \rk
\stackrel{k \to \infty}{\longrightarrow} 0
\ee
vanishes in the ultraviolet implying that the space of gauge orbits becomes
asymptotically flat in accordance with asymptotic freedom.

\subsection{Renormalization}

With the above obtained uv-behavior (see eqs. (\ref{1-62}), (\ref{1-68}) and
(\ref{1-71})) the integrals $I_d (k)$ (\ref{18a}) and $I_f (k)$ (\ref{19}) are
uv-convergent.
Thus, 
contrary to the $D = 3 + 1$ dimensional case in $D = 2 + 1$ the DSE for the
ghost and Coulomb form factors do not need
renormalization. What needs, however, renormalization is the curvature
$\chi (k)$ and the gap equation. The renormalization of these two equations is
carried out in exactly the same way as described in refs. \cite{Reinhardt:2007wh} and
\cite{R3}, i.e. basically by subtracting this equations at
a renormalization point $\mu$, which leads to the renormalized equations
\cite{Reinhardt:2007wh}
\be
\label{G95}
\chi (k) & = &
\chi (\mu) + \bar{\chi} (k) \hk , \hk \bar{\chi} (k) = I_\chi (k) -
I_\chi (\mu) \\
\label{G96}
\omega^2 (k) - \bar{\chi}^2 (k) & = & 
k^2 + \xi_0 + \Delta I^{(2)}_\omega (k) + 2
\bar{\chi} (k) \lk \xi + \Delta 
I^{(1)}_\omega (k) - \Delta I^{(1)}_\omega (0) \rk \hk .
\ee
Here 
$ \chi (\mu)$ and 
\be
\label{1-97}
\xi_0 & = & \omega^2 (\mu) - \mu^2 \nonumber\\
\xi & = & \chi (\mu) + I^{(1)}_\omega (0)
\ee
are finite renormalization constants. Furthermore, we have introduced the
abbreviation 
\be
\label{1-98}
\Delta I^{(n)}_\omega (k) = I^{(n)}_\omega (k) - I^{(n)}_\omega (\mu)
\ee
with
\be
\label{1-99}
I^{(n)}_\omega (k, \Lambda)  =  \frac{N_C}{2} \int^{\Lambda} \frac{d^2 q}{(2 \pi)^2}  
(\hat{{\bf k}} \hat{{\bf q}})^2 
\cdot \frac{d ({\bf k} -{\bf q})^2 f ({\bf k} - {\bf q})}{({\bf k} - {\bf q})^2} 
\cdot \frac{\left[\omega ({\bf q}) - \bar{\chi} ({\bf q})\right]^n - \left[\omega ({\bf k}) 
- \bar{\chi} ({\bf k})\right]^n }
{\omega ({\bf q})} \hk .
\ee
Note, since the ghost equation needs no renormalization, there are only three
renormalization constants
$ \xi_0$, 
$\xi$ and $\chi (\mu)$ and furthermore, the solutions of the coupled ghost and
gluon DSEs do not depend on $\chi (\mu)$\footnote{Some observables like the 't
Hooft loop (in $D = 3  + 1$) 
do, however, depend  on $\chi (\mu)$\cite{Reinhardt:2007wh}.}. (Note the integrals
(\ref{1-99}) depend only on the finite quantity $\bar{\chi} (k)$ but not on the
renormalization constant $\chi (\mu)$ \cite{Reinhardt:2007wh}.) 
One of the two independent
renormalization constants, $ \xi_0$, is used to implement the horizon condition
$d^{- 1} (k \to 0) = 0$. As will be shown in the next section
 any consistent solution
of the coupled DSEs does satisfy this condition.
The remaining renormalization constant $\xi$  can be chosen at will and
 determines the infrared limit of the wave functional as will be shown at the
 end of the next section.

\section{Analytic results}

The Dyson-Schwinger equations arising from the variational solution of the
Schr\"odinger equation have in principle the same form in $D = 2 + 1$ as in $D =
3 + 1$ dimensions. However, due to the fact that the $2 + 1$ dimensional theory
is
superrenormalizable some rigorous properties of the solution of the Schwinger-Dyson
equations can be derived which are not accessible in $3 + 1$ dimensions. Below
we shall derive some rigorous properties  of the solutions of the DSE and
discuss their physical implications.

\subsection{General results \label{sec4.1}}

Consider the ghost DSE (\ref{17a}). Since
 the integral $I_d (k)$ (\ref{18a}) is convergent in $D = 2 + 1$ (unlike in
$D = 3 + 1$), the ghost DSE needs no renormalization and thus no
renormalization constant is introduced by this equation. It is then not
difficult to prove the following statement: 
\bi

\no
If $d (k)$ is a continuous function
in $k \in [0, \infty)$, it satisfies 
\be
\label{an1}
d (k) \geq 1 \hk .
\ee
\bi

\no
We prove the statement by reductio ad absurdum: Assume $d (k) < 0$ for all $k
\in [0, \infty)$. Then, since $\omega (k) > 0$ by normalizability of the wave
functional it follows from (\ref{18a})  $I_d (k) \leq 0$ and thus from the
ghost DSE (\ref{17a})
\be
\label{an2}
\frac{1}{1 - I_d (k)} = d (k) > 0 \hk 
\ee
in contradiction to the assumption. Hence, $d (k)$ cannot be negative
everywhere. Assume now there exist some momentum $k'$ for which $d (k') < 0$.
Since $d (k)$ as a solution of the DSE, can be assumed to be continuous and 
as shown
above is not
everywhere negative, it must have at least one zero, say $k_0$, 
i.e. $d (k_0) = 0$, where $d (k)$ changes sign. By the ghost DSE, eq. (\ref{17a})
$I_d (k)$ has to be singular at $k \to k_0$ and change sign at $k = k_0$, too.
However, changing integration variable in the (convergent!) integral $I_d (k)$
(\ref{18a}) we find
\be
\label{an3}
I_d \lk k_0 \pm \epsilon k_0 \rk = \frac{N_C}{2} \int \frac{d^2 q}{(2 \pi)^2}
\lk 1 - \frac{\lk \vk_0 \cdot \lk \vq \pm \epsilon \vk_0 \rk \rk^2}{k^2_0 \lk
\vq \pm \epsilon \vk_0 \rk^2} \rk \frac{d \lk \vk_0 - \vq \rk}{\lk \vk_0 - \vq
\rk^2 \omega \lk \vq - \epsilon \vk_0 \rk} \hk 
\ee
and by the positivity of $\omega (k)$, $I_d \lk k_0 + \epsilon k_0 \rk$ and $I_d
\lk k_0 - \epsilon k_0 \rk$ have the same sign in contradiction to the above
assumption. Thus, $d (k) \geq 0$ holds for all $k$. Then from eq. (\ref{18a})
follows $I_d (k) \geq 0$ and by the ghost DSE (\ref{17a}) $d (k) = \lk 1 - I_d
(k) \rk^{- 1} \geq 1$. Note, that (\ref{an1}) is also in agreement with the
asymptotic uv-behavior found in the previous section. 

Assume now that the ghost form factor is bounded from above, i.e. there exist
some upperbound $M > 1$ such that $d (k) \leq M$ for $\forall k \in [0,
\infty)$. According to (\ref{an1}) $d (k)$ is then restricted to the intervall
$1 \leq d (k) \leq M$ and the integrand in $\chi (k)$ (\ref{22}) is positive
definite. Therefore, replacing $d (k)$ in the curvature (\ref{22}) by its upper
and lower bound, we obtain an upper and lower bound to $\chi (k)$
\begin{align}
M^2 I \geq \chi (k) \geq I \hk , \hk I = 
\frac{N_C}{2}  \il^{\Lambda} \frac {d^2 q}{(2 \pi)^2}
\left( 1 - \frac {({\bf k}{\bf q})^2}{{\bf k}^2 {\bf q}^2} \right)
\frac {1}{({\bf k} - {\bf q})^2} \hk ,
\label{2-211}
\end{align}
where we have introduced a momentum cutoff $\Lambda$ since the integral is
uv-divergent. The angular integral can be done analytically\footnote{Note that
this exact relation is also reproduced by the angular approximation.} 
\be
\label{11-44A}
\il^{\pi}_0 d \varphi \frac{\sin^2 \varphi}{k^2 + q^2 - 2 k q \cos \varphi} =
\frac{\pi}{2} \left\{ \begin{array}{ccc} \frac{1}{k^2} & , & q \leq k \\
\frac{1}{q^2} & , & k \leq q \end{array} \right. \hk 
\ee
yielding
\begin{align}
I = 
\frac{N_C}{8 \pi^2}  \left[ \il^{k}_0 dq q 
\frac{\pi}{k^2} + \il^{\Lambda}_k dq q \frac{\pi}{q^2} \right]  = \frac{N_C}{8
\pi} \left[ \frac{1}{2} + \ln \frac{\Lambda}{k} \right] .
\label{2-212}
\end{align}
Inserting this result into eq. (\ref{2-211}),
 we find for $\bar{\chi} (k) = \chi (k) - \chi (\mu)$ the bounds
\be
\label{11-47}
- \frac{N_C}{8 \pi} M^2 \ln \frac{k}{\mu} \geq \bar{\chi} (k) \geq -
\frac{N_C}{8 \pi} \ln \frac{k}{\mu} \hk .
\ee
Note that this relation holds for all $k$. In particular, this relation shows
that $\bar{\chi} (k)$ is infrared divergent
\be
\label{11-48}
\bar{\chi} (k) \sim - \ln \frac{k}{\mu} \hk , \hk k \to 0 \hk .
\ee
For an infrared divergent $\bar{\chi} (k)$ the gap equation (\ref{G95}) reduces
in the infrared to (see also ref. \cite{Reinhardt:2007wh})
\be
\label{11-49}
\lim\limits_{k \to 0} \omega (k) - \bar{\chi} (k) = \xi \hk ,
\ee
where $\xi$ is the renormalization constant introduced in eq. (\ref{1-97}). Eq.
(\ref{11-49}) shows that $\omega (k)$ has the same infrared behavior as
$\bar{\chi} (k)$. Thus, if the ghost form factor is bounded $1 \leq d (k) \leq M$
the gluon energy is logarithmically infrared divergent $\omega (k) \sim - \ln
\frac{k}{\mu}$. It is now not difficult to show that with such an infrared
behavior of $\omega (k)$ 
the ghost Dyson-Schwinger equation does not possess a solution. To
show this let us assume that there exist an $\epsilon > 0$ such that 
\be
\label{11-50}
\omega (k) \leq a \lk - \ln \frac{k}{\mu} \rk \hk , \hk \forall k \in (0,
\epsilon) \hk ,
\ee
with some positive constant $a$, which includes the case (\ref{11-48}).
Consider the integral in the ghost Dyson-Schwinger equation (\ref{18a})
\begin{align}
I_d (k) & =  \frac{N_C}{8 \pi^2}  \il^{\infty}_0 dq \frac{q}{\omega (q)} 
\il^{2 \pi}_0 d \varphi \sin^2 \varphi \cdot \frac {d({\bf k} - {\bf q})}{({\bf k} - {\bf q})^2}
\label{2-205}
\end{align}
Since $d (k) \geq1$, see eq. (\ref{an1}), we obtain the following estimate
\begin{align}
I_d (k) & \geq  \frac{N_C}{8 \pi^2}  \il^{\infty}_0 dq \frac{q}{\omega (q)} 
\il^{2 \pi}_0 d \varphi \sin^2 \varphi \cdot \frac {1}{({\bf k} - {\bf q})^2}
\geq \frac{N_C}{8 \pi^2}  \il^{\epsilon}_0 dq \frac{q}{\omega (q)} 
\il^{2 \pi}_0 d \varphi \sin^2 \varphi \cdot \frac {1}{({\bf k} - {\bf q})^2} .
\label{2-206}
\end{align}
The angular integral can be done exactly using eq. (\ref{11-44A}) resulting in 
\begin{align}
I_d (k) & \geq  \frac{N_C}{8 \pi^2}  \left[ \il^{k}_0 dq \frac{q}{\omega (q)} 
\frac{\pi}{k^2} + \il^{\epsilon}_k dq \frac{q}{\omega (q)} \frac{\pi}{q^2} \right]
\geq \frac{N_C}{8 \pi} \il^{\epsilon}_k dq \frac{1}{q \omega (q)} .
\label{2-207}
\end{align}
Inserting here eq. (\ref{11-50}) we obtain
\begin{align}
I_d (k) & \geq \frac{N_C}{8 \pi} \il^{\epsilon}_k dq \frac{1}{q a(- \ln
\frac{q}{\mu})}
= \frac{N_C}{8 \pi a} \ln \left| \frac{\ln \frac{k}{\mu}}{\ln
\frac{\epsilon}{\mu}} \right|
\hspace{0.25cm} \stackrel {k \to 0} {\longrightarrow} \hspace{0.25cm} \infty \hk
.
\label{2-208}
\end{align}
Thus  we find 
\begin{align}
I_d (k) & \hspace{0.25cm} \stackrel {k \to 0} {\longrightarrow} \hspace{0.25cm} \infty .
\label{2-209}
\end{align}
and from the ghost Dyson-Schwinger equation (\ref{17a}) follows $d (k \to 0) =
0$, which is in contradiction to the rigorous result (\ref{an1}). Thus we have
shown that the coupled ghost and gluon Dyson-Schwinger equations do not allow
for a ghost form factor which is bounded from above. We now show that ghost form
factor $d (k)$ is a monotonously decreasing function of $k$, i.e.
\be
\label{13X1}
d' (k) < 0
\ee
for all finite $k$.

As shown above in the ultraviolet analysis $d' (k) < 0$ for $k \to \infty$.
Assume now, as we lower $k$, at some finite $k = k_0$, $d' (k = k_0) = 0$. Then
from the eq. (\ref{1-56}) follows $I'_d (k_0) < 0$, which is in contradiction to
the ghost DSE from which follows
\be
\label{13X2}
\frac{d' (k)}{d^2 (k)} = I'_d (k) \hk .
\ee
Thus (\ref{13X1}) holds in the whole momentum range. Since (\ref{an1}) $d (k)
\geq 1$ and $d (k)$ is monotonously decreasing in the whole momentum range $k \geq
0$ and furthermore $d (k)$ must not be bounded from above, it follows that $d
(k)$ is infrared divergent, i.e. 
\be
d^{- 1} (k = 0) = 0 \hk ,
\ee
which is the horizon condition.

This is different 
from the $3 + 1$ dimensional case where solutions
to the Dyson-Schwinger equation exist with an infrared finite ghost form factor.

\subsection{Infrared analysis}

The DSE can be solved analytically in the infrared completely analogous to the
$D = 3 + 1$ dimensional case. For this purpose we make the following power
ans\"atze in the infrared
\be
\label{1-80}
\omega (k) = \frac{A}{k^\alpha} \hk , \hk d (k) = \frac{B}{k^\beta} \hk , \hk
\chi (k) = \frac{C}{k^\gamma} \hk .
\ee
We will first resort to the angular approximation used already in the
uv-analysis. Later on we will present the results obtained without resorting to
the angular approximation. 

With the infrared ans\"atze (\ref{1-80}), we find for the integrals defined by
eq. (\ref{1-59}) and (\ref{1-61})
\be
\label{1-79}
R (k \to 0) = \frac{1}{A} \frac{1}{1 + \alpha} k^{2 + \alpha} \hk , \hk S (k \to
0) = \frac{B}{1 - \beta} k^{2 - \beta} \hk .
\ee
From the derivative of the ghost DSE  (\ref{13X2})
we obtain
\be
\label{1-82}
\frac{A}{B^2} = \frac{N_C}{8 \pi} \frac{\beta + 2}{\beta (\alpha + 2)} k^{\alpha
- 2 \beta} \hk ,
\ee
which implies 
\be
\label{1-83}
\alpha = 2 \beta
\ee
and 
\be
\label{1-84}
\frac{A}{B^2} = \frac{N_C}{8 \pi} \frac{\beta + 2}{2\beta (\beta + 1)} \hk .
\ee
In an analogous fashion we obtain from the derivative of the curvature $\chi'
(k) = I'_\chi (k)$ 
\be
\label{1-86}
\frac{C}{B^2} = \frac{N_C}{8 \pi} \frac{\beta + 2}{\gamma (2 -\beta)} k^{\gamma
- 2 \beta}
\ee
resulting in 
\be
\label{1-87}
\gamma = 2 \beta
\ee
and
\be
\label{1-88}
\frac{C}{B^2} = \frac{N_C}{8 \pi} \frac{\beta + 2}{2 \beta (2 - \beta)} \hk .
\ee
From eq. (\ref{1-83}) and (\ref{1-87}) follows
\be
\label{1-89}
\alpha = \gamma = 2 \beta \hk ,
\ee
showing that $\omega (k)$ and $\chi (k)$ have the same infrared exponents just
like in the $D = 3 + 1$ dimensional case (ref. \cite{R3}). In fact, $\alpha =
\gamma$ follows already from the infrared limit of the gap equation
(\ref{11-49}). From this equation in  addition follows 
that not only the infrared exponents but also the prefactors of both
quantities have to coincide, i.e. 
\be
A = C \hk . 
\ee
Dividing eq. (\ref{1-84}) by
(\ref{1-88}) and using (\ref{1-89}) we obtain
\be
\label{1-90}
\frac{A}{C} = \frac{(2 - \beta)}{(1 + \beta)} \hk 
\ee
and $A = C$ 
\be
\label{1-91}
\beta = \frac{1}{2} \hk , \hk \alpha = \gamma = 1 \hk .
\ee
As we will see in sect. 6, this infrared behavior yields a linearly rising
 static
color Coulomb potential provided we approximate the Coulomb form factor $f (k)$
by its leading term $f (k \to 0) = 1$ which is correct to the order considered
in the present paper. 

The infrared analysis of the Dyson-Schwinger equations can be also
carried out without resorting to the angular approximation. In fact, in ref.
\cite{R7} the infrared analysis was carried out for arbitrary
dimensions. In that case one finds from the ghost DSE the following sum rule for
the infrared exponents
\be
\label{1-92}
\alpha = 2 \beta + d - 2 \hk 
\ee
due to the non-renormalization of the ghost-gluon vertex. 
As shown in ref. \cite{R7} this sum rule guarantees that $\chi (x)$
and $\omega (k)$ have the same infrared exponent $\alpha = \gamma$ as already
found above in the angular approximation, and in agreement with the infrared
limit of the gap equation (\ref{11-49}). Eq. (\ref{11-49}) together with the
ghost DSE in the infrared limit can be solved analytically for the infrared
exponents yielding \cite{R7}
\be
\label{1-94}
\beta = 0.4 \hspace{1cm}  i.e. \hspace{1cm}  \alpha = \gamma = 0.8  \hk ,
\ee
which is somewhat smaller than the infrared exponent found above in the angular
approximation (\ref{1-91}).

The above given infrared analysis is independent of the so far unfixed
renormalization constants $\xi$ and $\chi (\mu)$. (Recall that the coupled set
of DSEs (\ref{17a}), (\ref{G95}), (\ref{G96}) do not depend on $\chi (\mu)$.) From eq. (\ref{11-49}) 
we obtain $(\chi (k) =
\bar{\chi} (k) + \chi (\mu))$
\be
\lim\limits_{k \to 0} \lk \omega (k) - {\chi} (k) \rk = c \hk , \hk c = \xi -
\chi (\mu) \hk .
\ee
If one uses the infrared expressions for 
$\omega (k) , d (k), \chi(k) (\mbox{and} f (k) = 1)$ defined by eq.
(\ref{1-80}), one finds that the energy density is minimized for 
$c = 0$ (see ref. \cite{Reinhardt:2007wh}). Using the representation \cite{R4}
\be
\label{101}
\det J (a) = \exp \lk - \int A \chi A \rk \hk ,
\ee
which is correct to the order considered in the present paper, 
$c = 0$ implies an infrared limit of the wave functional
\be
\label{102}
\Psi (A) = \mbox{const} \pli_k \Psi (k) \hk , \hk \Psi (k \to 0) = 1 \hk .
\ee
This wave functional describes a stochastic vacuum, where the infrared modes of
the gauge field are completely unconstrained.

\section{Numerical results}

The coupled DSEs (\ref{22}), (\ref{17a}), (\ref{18}), (\ref{G96}) 
were solved numerically in the whole momentum range as
described in refs. \cite{R3} and \cite{R17}.
The renormalization constant $\xi_0$ was fixed by implementing the horizon
condition
$d^{- 1} (k = 0) = 0$. Furthermore, like in the $D = 3 + 1$ dimensional case in
order to stay consistently in 1-loop approximation we have solved the
equation for the Coulomb form factor $f (k)$ by assuming a bare ghost form factor
$d (k) = 1$ in the DSE for $f (k)$. 
\begin{figure} [!htb]
\label{fig2.14}
\begin{center}
\includegraphics [scale=0.61,bb=47 29 601 436,clip=] {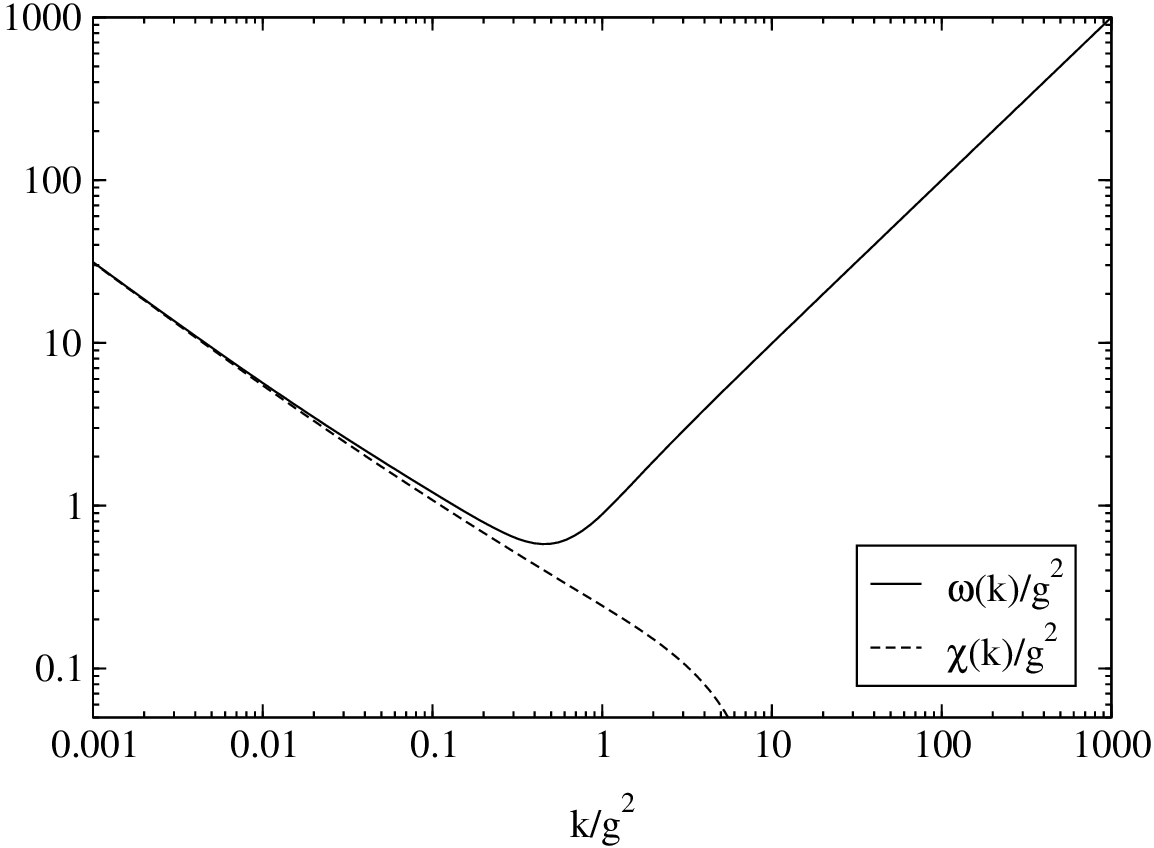}
\caption{ The gluon energy $\omega (k)$ and the curvature $\chi (k)$ obtained from the 
numerical solution of the DSEs for $\xi = 2.0 g^2$ and $\chi (\mu) = 0.5 g^2 $.}
\end{center}
\end{figure}
\begin{figure} [!htb]
\begin{center}\label{fig2.15}
\includegraphics [scale=0.61,bb=66 29 601 430,clip=] {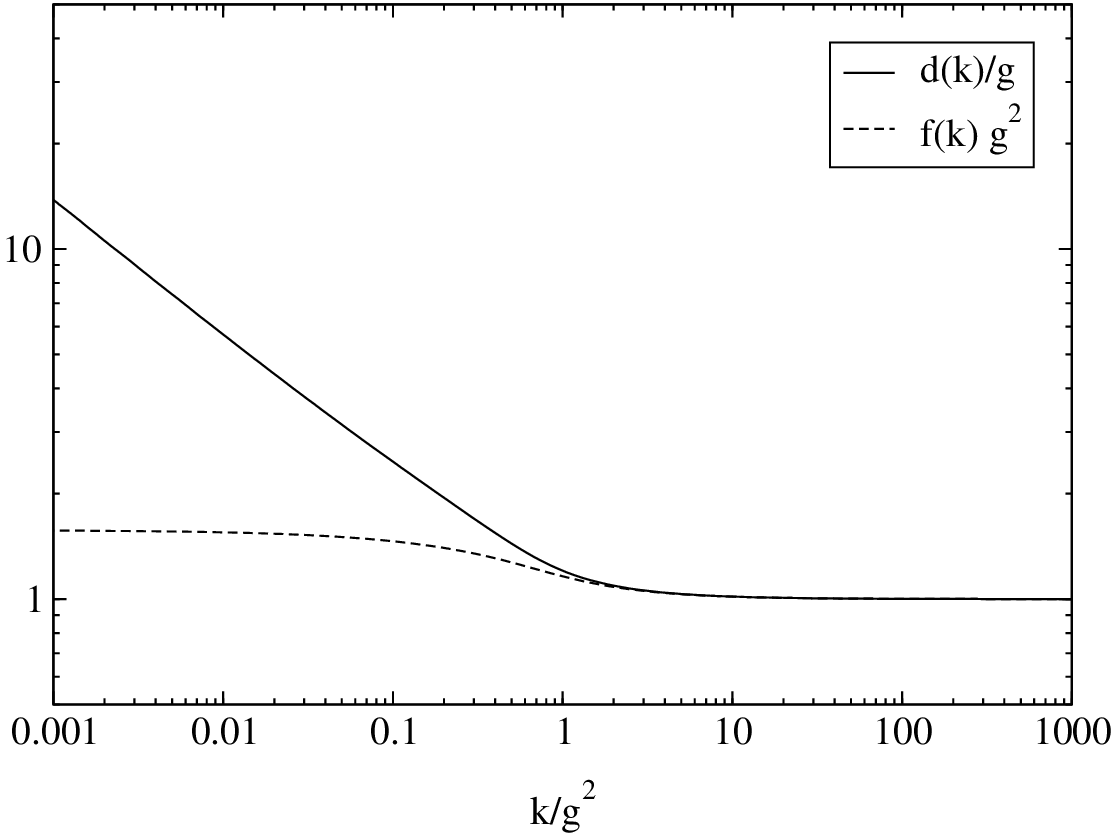}
\caption{ 
The ghost form factor $d(k)$ and the Coulomb form factor $f(k)$
for $\xi = 2.0 g^2$ and $\chi (\mu) = 0.5 g^2 $.}
\end{center}
\end{figure}
\begin{figure} [!htb]
\begin{center}
\includegraphics [scale=0.61,bb=28 28 588 433,clip=] {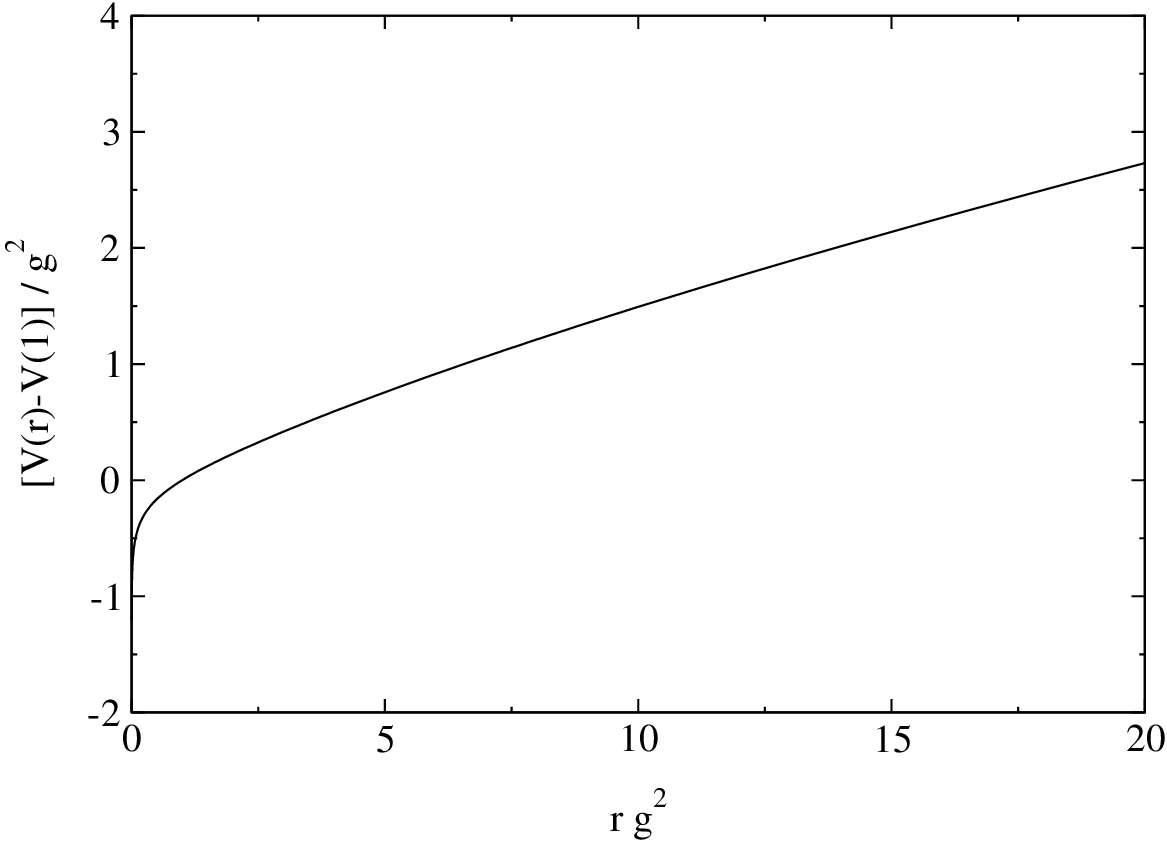}
\caption{ 
The static non-Abelian Coulomb potential for
$\xi = 2.0 g^2$ and $\lambda (\mu) = 0.5 g^2 $.}
\end{center}
\end{figure}
The numerical results obtained are presented in figs. 1 and 2. 
Fig. 1 shows the gluon energy 
$\omega (k)$ and the curvature
$\chi (k)$. Both quantities are infrared divergent and approach each other for 
$k \to 0$ in agreement with our infrared analysis given in the previous section.
Fig. 2 shows the ghost and the Coulomb form factor. The ghost form
factor is of course infrared divergent, since  as shown in sect. 4, 
self-consistent solution of the Dyson-Schwinger
equations exist only for infrared divergent ghost form factors. The Coulomb
form factor is infrared finite and approaches asymptotically for 
$k \to \infty$ the ghost form factor.
\begin{figure} [!htb]
\begin{center}
\includegraphics [scale=0.49,bb=34 30 526 528,clip=] {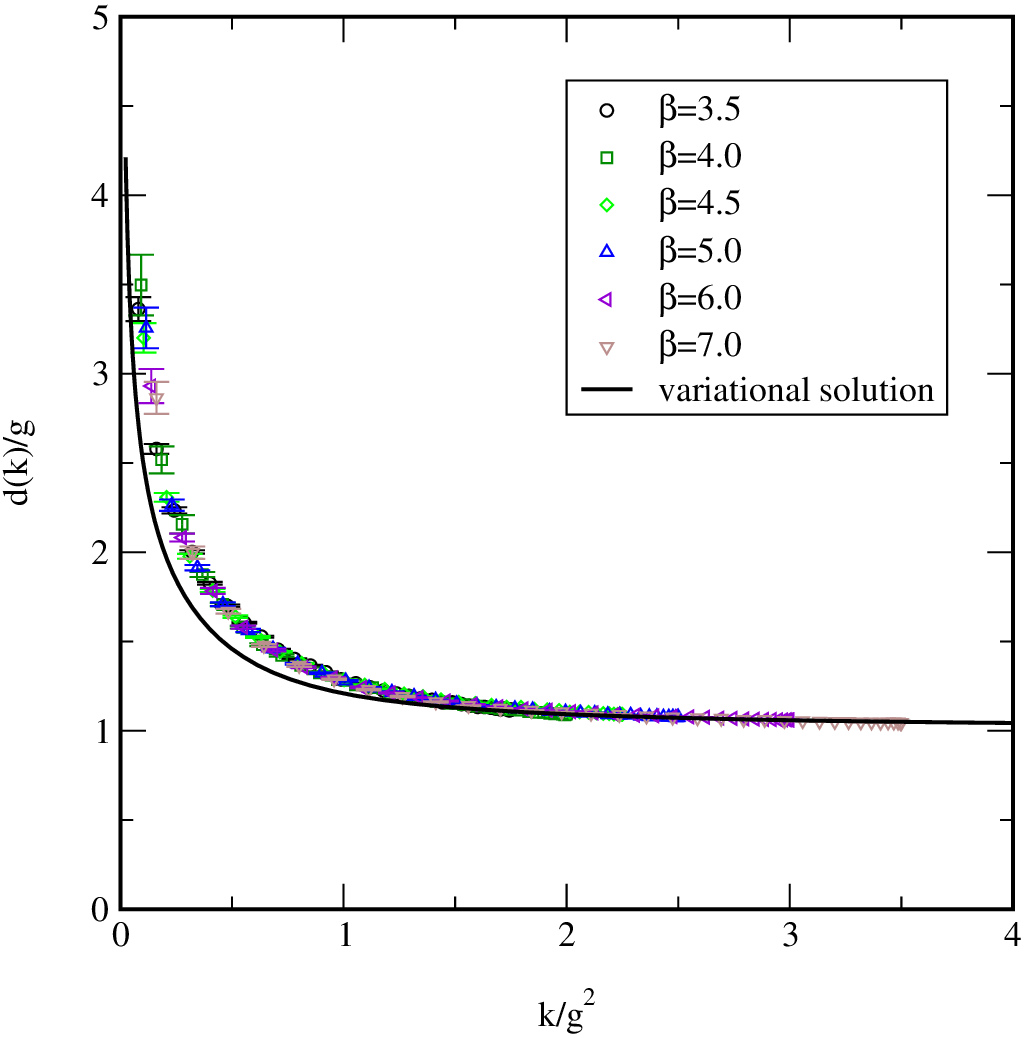}
\caption{ 
Comparison of the ghost form factor $d(k)$ obtained from the numerical solution of the DSEs for 
$\xi = 2.0 g^2$ and $\chi (\mu) = 0.5 g^2 $ with the lattice data obtained in \cite{R8}.}
\end{center}
\end{figure}
\begin{figure} [!htb]
\begin{center}
\includegraphics [scale=0.49,bb=30 30 525 528,clip=] {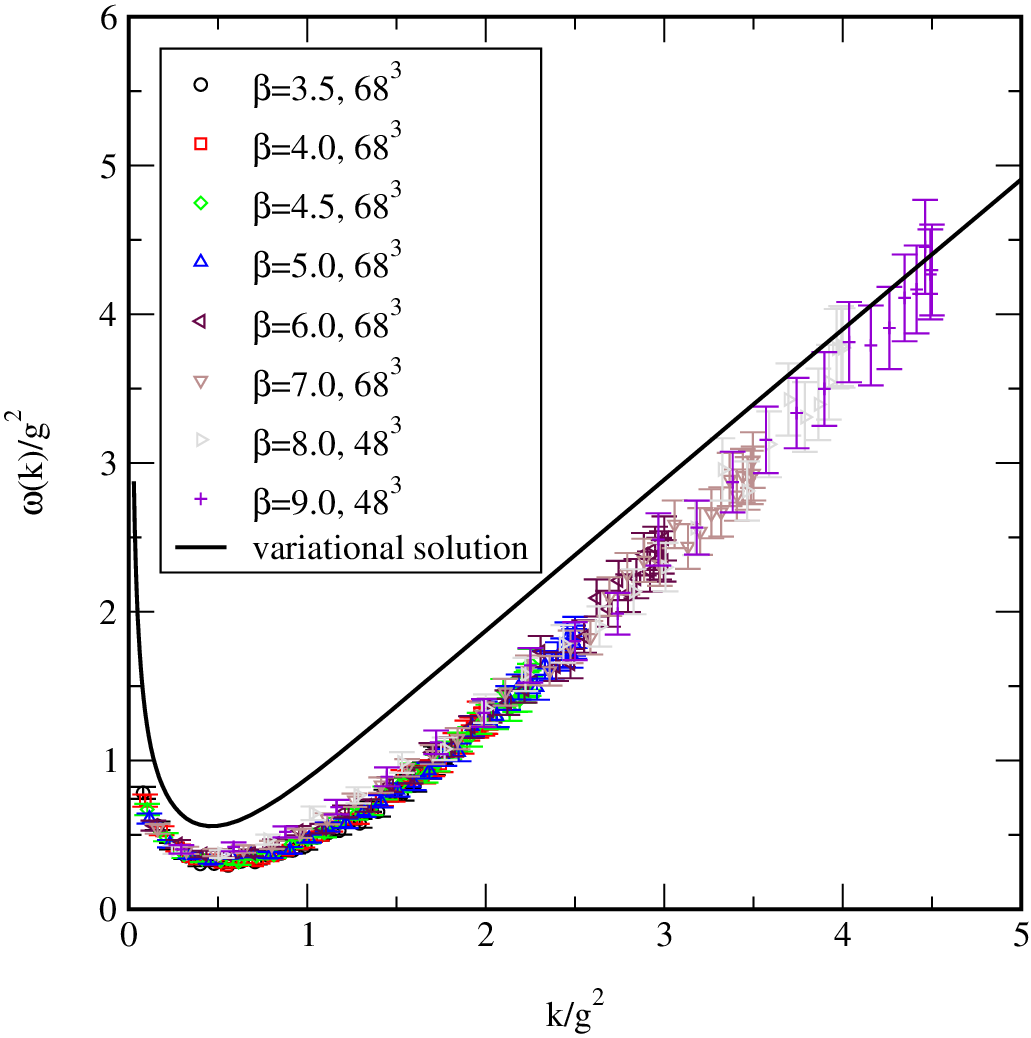}
\caption{ 
The gluon energy $\omega (k)$ obtained from the numerical solutions of the DSEs for
$\xi = 2.0 g^2$ and $\chi (\mu) = 0.5 g^2 $ and the corresponding lattice data
obtained in \cite{R8}.}
\end{center}
\end{figure}
\begin{figure} [!htb]
\begin{center}
\includegraphics [scale=0.49,bb=31 30 526 528,clip=] {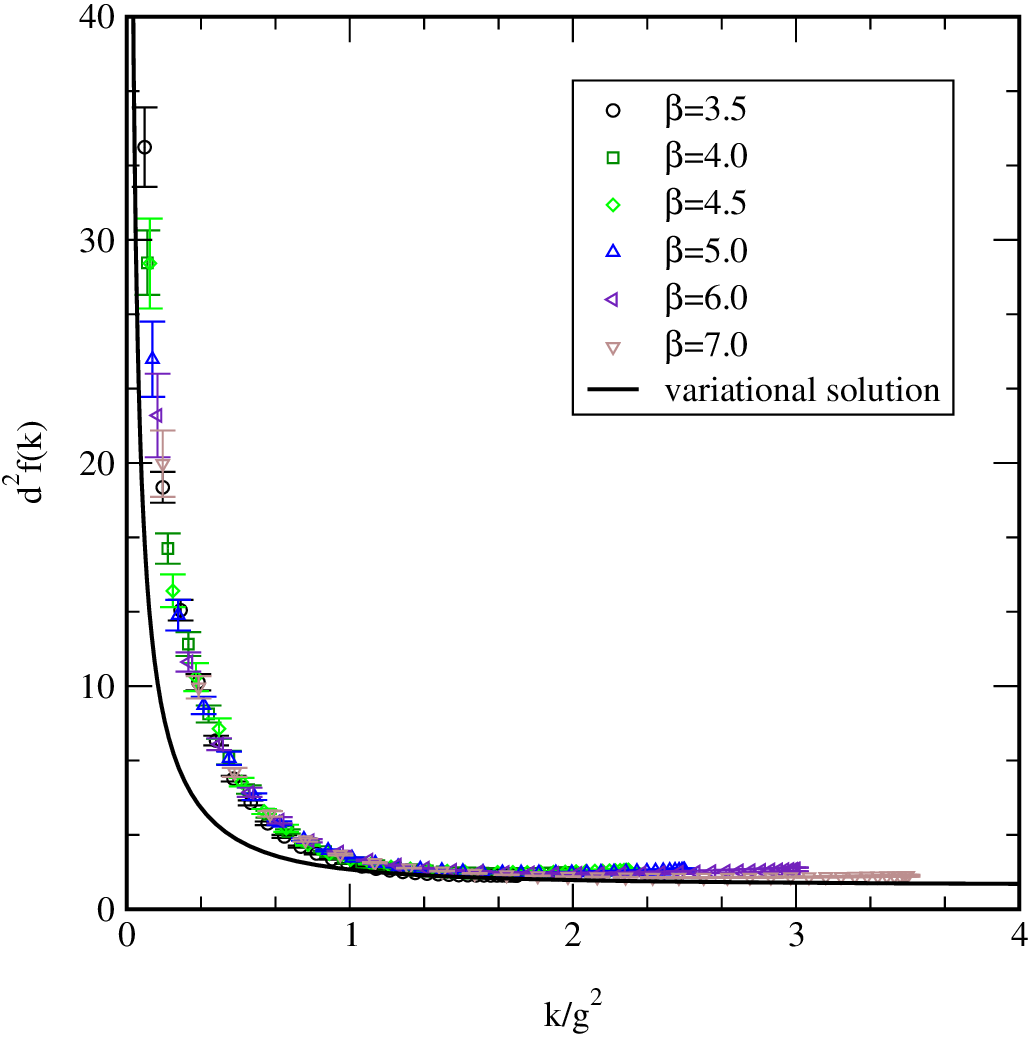}
\caption{ 
The Fourier transform of the static potential obtained from the numerical solution of the
DSEs and the corresponding lattice data obtained in \cite{R8}.}
\end{center}
\end{figure}
Fig. 3 shows the static color Coulomb potential defined by \cite{R3}
\be
\label{1-103}
V \lk {\bf r} \rk =  \int \frac{d^2 k}{(2 \pi)^2}  V (k)
e^{i{\bf k} {\bf r}} =  \int \frac{d^2 k}{(2 \pi)^2} \frac{ d (k)^2 f
(k)}{k^2} e^{i {\bf k} {\bf r}}  = \frac{1}{2 \pi} \il^{\infty}_0 dk \frac{ d (k)^2 f
(k)}{k^2} J_0 (kr) \hk ,
\ee
where 
$J_0 (k)$ is the (ordinary) zero'th order Bessel function. For the infrared
behavior obtained in the angular approximation
$ \beta = 1/2$, see eq. (\ref{1-80}), 
$V (k) \sim \raisebox{1mm}{\small 1} \hspace{ -1mm }
\raisebox{0mm}{\small /} \hspace{ -1mm }
\raisebox{-1mm}{\small $ k^3$}$, which leads to a strictly linearly rising quark
potential at large distances, while for
$k \to \infty$, where
$d (k \to \infty) = f (k \to \infty) = 1$ the potential behaves as
$V (k) \sim \raisebox{1mm}{\small 1} \hspace{ -1mm }
\raisebox{0mm}{\small /} \hspace{ -1mm }
\raisebox{-1mm}{\small $k^2$}$ and we obtain the familiar Coulomb potential
$V (r) \sim \ln \raisebox{1mm}{\small r} \hspace{ -1mm }
\raisebox{0mm}{\small /} \hspace{ -1mm }
\raisebox{-1mm}{\small $r_0$}$
in $D = 2 + 1$ dimensions. The infrared analysis carried out without resorting
to the angular approximation yields
$V (k) \sim \raisebox{1mm}{\small 1} \hspace{ -1mm }
\raisebox{0mm}{\small /} \hspace{ -1mm }
\raisebox{-1mm}{\small $k^{2.8}$} , k \to 0$. A careful analysis of our numerical
solutions (obtained without the angular approximation) yields
$V (k) \sim \raisebox{1mm}{\small 1} \hspace{ -1mm }
\raisebox{0mm}{\small /} \hspace{ -1mm }
\raisebox{-1mm}{\small $k^{2.9}$} , k \to 0$, which is in between the analytical
results obtained with and without the angular approximation. 
In figs. 4, 5 and 6 we compare our numerical results for the
ghost form factor
$d (k)$, the gluon energy $\omega (k)$ and the form factor of the Coulomb
potential $d^2 (k) f (k)$ with the lattice data. It has been shown by solving
the Dyson-Schwinger equation in Landau gauge on the torus \cite{Fischer:2002eq},
 \cite{Fischer:2007pf},
that very large lattices are required to capture the
correct infrared behavior of the Greens functions of the continuum theory.
While these lattice sizes can be reached in 3 dimensions, they are out of reach
in $D = 4$. The lattice calculations performed in ref. \cite{R8}
in $2 + 1$ dimensions used lattices of the size $64^3$, which should be
sufficient to extract the correct infrared limit of the Greens functions. Our
numerical results obtained by solving the DSE in Coulomb gauge are in quite
satisfactory agreement with the lattice data. In particular, the asymptotic
ultraviolet and infrared behaviors are quite well reproduced.

\section{Summary and Conclusions}

We have performed a variational solution of the Yang-Mills Schr\"odinger equation
in Coulomb gauge in $D = 2 + 1$. The Dyson-Schwinger equations resulting from the
minimization of the vacuum energy density have been solved analytically in
the ultraviolet and in the infrared and in addition some rigorous results of
their properties have been derived. In particular, we have shown that the ghost
form factor as well as the gluon energy have to be infrared divergent, which is
different from the $3 + 1$ dimensional case where solutions of the DSEs exist
with these quantities being infrared finite \cite{rsx}.  The static non-Abelian Coulomb
potential resulting from our numerical solution of the DSE is almost linearly
rising. Our numerical results are in satisfactory agreement with the existing
lattice data.
The lattice calculations performed in $D = 3 + 1$ so far use too
small lattices  
to give reliable results for the continuum
limit, in particular, on the infrared properties of the various Green's
functions \cite{Fischer:2002eq}, \cite{Fischer:2007pf}.
\bi

\no
{\Large \bf Acknowledgments}
\bi

\no
Discussions with G. Burgio, D. Epple, M. Quandt, W. Schleifenbaum, A. Szczepaniak
and A. Weber are gratefully acknowledged.


\begin{thebibliography}{99}
\bibitem{R1}
  D.~Zwanziger,
  Phys.\ Rev.\  D {\bf 70} (2004) 094034
  [arXiv:hep-ph/0312254].
\bibitem{R2}
  A.~P.~Szczepaniak and E.~S.~Swanson,
  Phys.\ Rev.\ D {\bf 65} (2001) 025012
  [arXiv:hep-ph/0107078];
  A.~P.~Szczepaniak,
  Phys.\ Rev.\ D {\bf 69} (2004) 074031
  [arXiv:hep-ph/0306030].
\bibitem{R3}
  C.~Feuchter and H.~Reinhardt,
  Phys.\ Rev.\ D {\bf 70} (2004) 105021
  [arXiv:hep-th/0408236], 
  arXiv:hep-th/0402106.
\bibitem{R4}
  H.~Reinhardt and C.~Feuchter,
  Phys.\ Rev.\ D {\bf 71} (2005) 105002
  [arXiv:hep-th/0408237].
\bibitem{R5}
  J.~Greensite, S.~Olejnik and D.~Zwanziger,
  Phys.\ Rev.\  D {\bf 69} (2004) 074506
  [arXiv:hep-lat/0401003].
\bibitem{R6}
  K.~Langfeld and L.~Moyaerts,
  Phys.\ Rev.\ D {\bf 70} (2004) 074507
  [arXiv:hep-lat/0406024].
\bibitem{R6a}
  S.~Furui and H.~Nakajima,
  arXiv:0708.1421 [hep-lat].
\bibitem{R6b}
  A.~Cucchieri and D.~Zwanziger,
  Phys.\ Rev.\  D {\bf 65} (2002) 014001
  [arXiv:hep-lat/0008026].
\bibitem{R8}
L. Moyaerts, PhD thesis, University of T\"ubingen.
\bibitem{R10}
N. H. Christ and T. D. Lee, Phys. Rev. {\bf D22} (1980) 939; Phys. Scripta {\bf 23}
(1981) 970.
\bibitem{R7}
  W.~Schleifenbaum, M.~Leder and H.~Reinhardt,
  Phys.\ Rev.\ D {\bf 73} (2006) 125019
  [arXiv:hep-th/0605115].
\bibitem{Fischer:2002eq}
  C.~S.~Fischer, R.~Alkofer and H.~Reinhardt,
  Phys.\ Rev.\  D {\bf 65}, 094008 (2002)
  [arXiv:hep-ph/0202195].
\bibitem{Fischer:2007pf}
  C.~S.~Fischer, A.~Maas, J.~M.~Pawlowski and L.~von Smekal,
  arXiv:hep-ph/0701050.
\bibitem{R11}
K. Johnson, The Yang-Mills Ground State, Proceedings of the workshop ``QCD - 20
Years Later'', Aachen, 1992, edited by P.M. Zerwas and H.A. Kastrup, Vol 7; D.Z.
Freedman, P.E. Haagensen, K. Johnson, and J.-I. Latorre, hep-th/9309045; N,
Bazer, D.Z. Freedman, and P.E. Haagensen, Nucl. Phys. {\bf B428}, 147 (1994)        
\bibitem{R12}
       I.L. Kogan and A. Kovner, Phys. Rev. {\bf D 52}, 3719 (1995); C. Heinemann,
       C. Martin, E. Jancu and D. Vautherin, Phys. Rev. {\bf D 61} 116008
       (2000); O. Schr\"oder and H. Reinhardt, Ann. Phys. (N.Y.) {\bf 312} 319
       (2004); O. Schr\"oder and H. Reinhardt, Ann. Phys. (N.Y.) {\bf 307} 452
       (2003)
\bibitem{Rx}
       J. Greensite, S. Olejnik, arXiv:0707.2860 [hep-lat] 
\bibitem{R13} 
        {J.C. Taylor, Nucl. Phys. {\bf B33} (1971) 436}
\bibitem{Maas:2007uv}
  A.~Maas,
  Phys.\ Rev.\  D {\bf 75}, 116004 (2007)
  [arXiv:0704.0722 [hep-lat]].
\bibitem{R15}
A. R. Swift, Phys. Rev. {\bf D38} (1988) 668
\bibitem{Reinhardt:2007wh}
  H.~Reinhardt and D.~Epple,
  Phys. Rev. {\bf D76} (2007) 065015,
  arXiv:0706.0175 [hep-th].
\bibitem{R17}
  D.~Epple, H.~Reinhardt and W.~Schleifenbaum,
  Phys.\ Rev.\  D {\bf 75} (2007) 045011
  [arXiv:hep-th/0612241].
\bibitem{'t Hooft:1977hy}
  G.~'t Hooft,
  Nucl.\ Phys.\  B {\bf 138}, 1 (1978).

\bibitem{Reinhardt:2002mb}
  H.~Reinhardt,
  Phys.\ Lett.\  B {\bf 557}, 317 (2003)
  [arXiv:hep-th/0212264].
\bibitem{rsx}
 {D. Epple, H. Reinhardt, W. Schleifenbaum and A. Szczepaniak, in preparation}
\end{thebibliography}
\end{document}